\documentstyle[epsfig,a4p,12pt,rotating]{article}
\begin{document}
%=======================================================================
%       Parameters for the title page
%=======================================================================
%
%  PPE number, Date and Version
%
\newcommand{\PPEnum}    {CERN-EP/98-039}
\newcommand{\PNnum}     {OPAL Physics Note PN-306}
\newcommand{\TNnum}     {OPAL Technical Note TN-xxx}
\newcommand{\Date}      {6th March 1998}
\newcommand{\Author}    {M.~Fanti and P.~Giacomelli}
\newcommand{\MailAddr}  {marcello.fanti@cern.ch or paolo.giacomelli@cern.ch}
\newcommand{\EdBoard}   {T.~Junk, H.~Neal, M.~Thomson and S.~Towers}
\newcommand{\DraftVer}  {Version 2.0}
\newcommand{\DraftDate} {6th March 1998}
\newcommand{\TimeLimit} {February 23th, 9:00}

%=======================================================================
%       Parameters affecting the general appearance
%=======================================================================
\def\toprule{\noalign{\hrule \medskip}}
\def\midrule{\noalign{\medskip\hrule }}
\def\botrule{\noalign{\medskip\hrule }}
\setlength{\parskip}{\medskipamount}

%=======================================================================
%       Define symbols
%=======================================================================

\newcommand{\LL}{{\mathrm L}^+ {\mathrm L}^-}
\newcommand{\QQ}{{\mathrm Q}\bar{\mathrm Q}}
\newcommand{\XX}{{\mathrm X}^+ {\mathrm X}^-}
\newcommand{\ee}{{\mathrm e}^+ {\mathrm e}^-}
\newcommand{\sq}{\tilde{\mathrm q}}
\newcommand{\seff}{\tilde{\mathrm f}}
\newcommand{\sele}{\tilde{\mathrm e}}
\newcommand{\sell}{\tilde{\ell}}
\newcommand{\snu}{\tilde{\nu}}
\newcommand{\supq}{\tilde{u}}
\newcommand{\sdown}{\tilde{d}}
\newcommand{\smu}{\tilde{\mu}}
\newcommand{\stau}{\tilde{\tau}}
\newcommand{\chp}{\tilde{\chi}^+_1}
\newcommand{\chn}{\tilde{\chi}^+_n}
\newcommand{\chpm}{\tilde{\chi}^\pm_1}
\newcommand{\nt}{\tilde{\chi}^0}
\newcommand{\qq}{{\mathrm q}\bar{\mathrm q}}
\newcommand{\sleppair}{\sell^+ \sell^-}
\newcommand{\nunu}{\nu \bar{\nu}}
\newcommand{\mumu}{\mu^+ \mu^-}
\newcommand{\tautau}{\tau^+ \tau^-}
\newcommand{\ellell}{\ell^+ \ell^-}
\newcommand{\nulqq}{\nu \ell {\mathrm q} \bar{\mathrm q}'}
\newcommand{\MZ}{M_{\mathrm Z}}
\newcommand{\eeff}{\ee\mathrm{f\bar{f}}}
\newcommand{\msl}{$m_{\sell}$}

\newcommand {\stopm}         {\tilde{\mathrm{t}}_{1}}
\newcommand {\stops}         {\tilde{\mathrm{t}}_{2}}
\newcommand {\stopbar}       {\bar{\tilde{\mathrm{t}}}_{1}}
\newcommand {\stopx}         {\tilde{\mathrm{t}}}
\newcommand {\sneutrino}     {\tilde{\nu}}
\newcommand {\slepton}       {\tilde{\ell}}
\newcommand {\stopl}         {\tilde{\mathrm{t}}_{\mathrm L}}
\newcommand {\stopr}         {\tilde{\mathrm{t}}_{\mathrm R}}
\newcommand {\stoppair}      {\tilde{\mathrm{t}}_{1}
\bar{\tilde{\mathrm{t}}}_{1}}
\newcommand {\gluino}        {\tilde{\mathrm g}}

\newcommand {\neutralino}    {\tilde{\chi }^{0}_{1}}
\newcommand {\neutrala}      {\tilde{\chi }^{0}_{2}}
\newcommand {\neutralb}      {\tilde{\chi }^{0}_{3}}
\newcommand {\neutralc}      {\tilde{\chi }^{0}_{4}}
\newcommand {\bino}          {\tilde{\mathrm B}^{0}}
\newcommand {\wino}          {\tilde{\mathrm W}^{0}}
\newcommand {\higginoa}      {\tilde{\rm H_{1}}^{0}}
\newcommand {\higginob}      {\tilde{\mathrm H_{1}}^{0}}
\newcommand {\chargino}      {\tilde{\chi }^{\pm}_{1}}
\newcommand {\chargib}       {\tilde{\chi }^{\pm}_{2}}
\newcommand {\charginop}     {\tilde{\chi }^{+}_{1}}
\newcommand {\chargibp}      {\tilde{\chi }^{+}_{2}}
\newcommand {\KK}            {{\mathrm K}^{0}-\bar{\mathrm K}^{0}}
\newcommand {\ff}            {{\mathrm f} \bar{\mathrm f}}
\newcommand {\bstopm} {\mbox{$\boldmath {\tilde{\mathrm{t}}_{1}} $}}
\newcommand {\Mt}            {M_{\mathrm t}}
\newcommand {\mscalar}       {m_{0}}
\newcommand {\Mgaugino}      {M_{1/2}}
\newcommand {\rs}            {\sqrt{s}}
\newcommand {\WW}            {{\mathrm W}^+{\mathrm W}^-}
\newcommand {\MGUT}          {M_{\mathrm {GUT}}}
\newcommand {\Zboson}        {{\mathrm Z}^{0}}
\newcommand {\Wpm}           {{\mathrm W}^{\pm}}
\newcommand {\allqq}         {\sum_{q \neq t} q \bar{q}}
\newcommand {\mixang}        {\theta _{\mathrm {mix}}}
\newcommand {\thacop}        {\theta _{\mathrm {Acop}}}
\newcommand {\cosjet}        {\cos\thejet}
\newcommand {\costhr}        {\cos\thethr}
\newcommand {\djoin}         {d_{\mathrm{join}}}
\newcommand {\mstop}         {m_{\stopm}}
\newcommand {\msell}         {m_{\sell}}
\newcommand {\mchi}          {m_{\neutralino}}
\newcommand {\pp}{p \bar{p}}

\newcommand{\epair}{\mbox{${\mathrm e}^+{\mathrm e}^-$}}
\newcommand{\mupair}{\mbox{$\mu^+\mu^-$}}
\newcommand{\taupair}{\mbox{$\tau^+\tau^-$}}
\newcommand{\qpair}{\mbox{${\mathrm q}\overline{\mathrm q}$}}
\newcommand{\eeee}{\mbox{\epair\epair}}
\newcommand{\eemumu}{\mbox{\epair\mupair}}
\newcommand{\eetautau}{\mbox{\epair\taupair}}
\newcommand{\eeqq}{\mbox{\epair\qpair}}
\newcommand{\fs}{ final states}
\newcommand{\epairf}{\mbox{\epair\fs}}
\newcommand{\mupairf}{\mbox{\mupair\fs}}
\newcommand{\taupairf}{\mbox{\taupair\fs}}
\newcommand{\qpairf}{\mbox{\qpair\fs}}
\newcommand{\eeeef}{\mbox{\eeee\fs}}
\newcommand{\eemumuf}{\mbox{\eemumu\fs}}
\newcommand{\eetautauf}{\mbox{\eetautau\fs}}
\newcommand{\eeqqf}{\mbox{\eeqq\fs}}
\newcommand{\ffff}{four fermion final states}
\newcommand{\llnunu}{\mbox{\lpair\nul\nubar}}
\newcommand{\lnuqq}{\mbox{\lept\nubar\qpair}}
\newcommand{\zee}{\mbox{Zee}}
\newcommand{\zzg}{\mbox{ZZ/Z$\gamma$}}
\newcommand{\wenu}{\mbox{We$\nu$}}

\newcommand{\el}{\mbox{${\mathrm e}^-$}}
\newcommand{\selem}{\mbox{$\tilde{\mathrm e}^-$}}
\newcommand{\smum}{\mbox{$\tilde\mu^-$}}
\newcommand{\staum}{\mbox{$\tilde\tau^-$}}
\newcommand{\slept}{\mbox{$\tilde{\ell}^\pm$}}
\newcommand{\sleptm}{\mbox{$\tilde{\ell}^-$}}
\newcommand{\lept}{\mbox{$\ell^-$}}
\newcommand{\Hl}{\mbox{$\mathrm{L}^\pm$}}
\newcommand{\Hm}{\mbox{$\mathrm{L}^-$}}
\newcommand{\Hnu}{\mbox{$\nu_{\mathrm{L}}$}}
\newcommand{\nul}{\mbox{$\nu_\ell$}}
\newcommand{\nubar}{\mbox{$\overline{\nu}_\ell$}}
\newcommand{\spair}{\mbox{$\tilde{\ell}^+\tilde{\ell}^-$}}
\newcommand{\lpair}{\mbox{$\ell^+\ell^-$}}
\newcommand{\staupair}{\mbox{$\tilde{\tau}^+\tilde{\tau}^-$}}
\newcommand{\smupair}{\mbox{$\tilde{\mu}^+\tilde{\mu}^-$}}
\newcommand{\selepair}{\mbox{$\tilde{\mathrm e}^+\tilde{\mathrm e}^-$}}
\newcommand{\ch}{\mbox{$\tilde{\chi}^\pm_1$}}
\newcommand{\chpair}{\mbox{$\tilde{\chi}^+_1\tilde{\chi}^-_1$}}
\newcommand{\chm}{\mbox{$\tilde{\chi}^-_1$}}
\newcommand{\chmp}{\mbox{$\tilde{\chi}^\pm_1$}}
\newcommand{\chz}{\mbox{$\tilde{\chi}^0_1$}}
\newcommand{\dch}{\mbox{\chm$\rightarrow$\chz\lept\nubar}}
\newcommand{\dslept}{\mbox{\sleptm$\rightarrow$\chz\lept}}
\newcommand{\dH}{\mbox{\Hm$\rightarrow$\lept\nubar\Hnu}}
\newcommand{\mch}{\mbox{$m_{\tilde{\chi}^\pm_1}$}}
\newcommand{\mslept}{\mbox{$m_{\tilde{\ell}}$}}
\newcommand{\mstau}{\mbox{$m_{\staum}$}}
\newcommand{\msmu}{\mbox{$m_{\smum}$}}
\newcommand{\msele}{\mbox{$m_{\selem}$}}
\newcommand{\mchz}{\mbox{$m_{\tilde{\chi}^0_1}$}}
\newcommand{\dm}{\mbox{$\Delta m$}}
\newcommand{\dmch}{\mbox{$\Delta m_{\ch-\chz}$}}
\newcommand{\dmslept}{\mbox{$\Delta m_{\slept-\chz}$}}
\newcommand{\dmhl}{\mbox{$\Delta m_{\Hl-\Hnu}$}}
\newcommand{\w}{\mbox{W$^\pm$}}

\newcommand{\acopc}{\mbox{$\phi^{\mathrm{acop}}$}}
\newcommand{\acolc}{\mbox{$\theta^{\mathrm{acol}}$}}
\newcommand{\acop}{\mbox{$\phi_{\mathrm{acop}}$}}
\newcommand{\acol}{\mbox{$\theta_{\mathrm{acol}}$}}
\newcommand{\pt}{\mbox{$p_{t}$}}
\newcommand{\pz}{\mbox{$p_{\mathrm{z}}^{\mathrm{miss}}$}}
\newcommand{\ptevt}{\mbox{$p_{t}^{\mathrm{miss}}$}}
\newcommand{\ptaxic}{\mbox{$a_{t}^{\mathrm{miss}}$}}
\newcommand{\stevt}{\mbox{$p_{t}^{\mathrm{miss}}$/\Ebeam}}
\newcommand{\staxic}{\mbox{$a_{t}^{\mathrm{miss}}$/\Ebeam}}
\newcommand{\dptaxic}{\mbox{missing $p_{t}$ wrt. event axis \ptaxic}}
\newcommand{\cosevt}{\mbox{$\mid\cos\theta_{\mathrm{p}}^{\mathrm{miss}}\mid$}}
\newcommand{\axicos}{\mbox{$\mid\cos\theta_{\mathrm{a}}^{\mathrm{miss}}\mid$}}
\newcommand{\pthet}{\mbox{$\theta_{\mathrm{p}}^{\mathrm{miss}}$}}
\newcommand{\athet}{\mbox{$\theta_{\mathrm{a}}^{\mathrm{miss}}$}}
\newcommand{\dcosevt}{\mbox{$\mid\cos\theta\mid$ of missing p$_{t}$}}
\newcommand{\daxicos}{\mbox{$\mid\cos\theta\mid$ of missing p$_{t}$ wrt. event
axis}}
\newcommand{\efdsw}{\mbox{$x_{\mathrm{FDSW}}$}}
\newcommand{\acopf}{\mbox{$\Delta\phi_{\mathrm{FDSW}}$}}
\newcommand{\acopm}{\mbox{$\Delta\phi_{\mathrm{MUON}}$}}
\newcommand{\acopt}{\mbox{$\Delta\phi_{\mathrm{trk}}$}}
\newcommand{\po}{\mbox{$E_{\mathrm{isol}}^\gamma$}}
\newcommand{\qprod}{\mbox{$q1$$*$$q2$}}
\newcommand{\lcode}{lepton identification code}
\newcommand{\nctro}{\mbox{$N_{\mathrm{trk}}^{\mathrm{out}}$}}
\newcommand{\necao}{\mbox{$N_{\mathrm{ecal}}^{\mathrm{out}}$}}
\newcommand{\mout}{\mbox{$m^{\mathrm{out}}$}}
\newcommand{\nctec}{\mbox{\nctro$+$\necao}}
\newcommand{\gfract}{\mbox{$F_{\mathrm{good}}$}}
\newcommand{\zz}       {\mbox{$|z_0|$}}
\newcommand{\dz}       {\mbox{$|d_0|$}}
\newcommand{\sint}      {\mbox{$\sin\theta$}}
\newcommand{\cost}      {\mbox{$\cos\theta$}}
\newcommand{\mcost}     {\mbox{$|\cos\theta|$}}
\newcommand{\dedx}     {{\mathrm d}E/{\mathrm d}x}
\newcommand{\wdedx}     {\mbox{$W_{dE/dx}$}}
\newcommand{\xe}     {\mbox{$x_E$}}

\newcommand{\ssix}     {\mbox{$\sqrt{s}$~=~161~GeV}}
\newcommand{\sthree}     {\mbox{$\sqrt{s}$~=~130--136~GeV}}
\newcommand{\mrecoil}     {\mbox{$m_{\mathrm{recoil}}$}}
\newcommand{\llmass}     {\mbox{$m_{ll}$}}
\newcommand{\sml}{\mbox{Standard Model \lpair$\nu\nu$ events}}
\newcommand{\sme}{\mbox{Standard Model events}}
\newcommand{\sig}
  {events containing a lepton pair plus missing transverse momentum}
\newcommand{\wpair}{\mbox{$W^+W^-$}}
\newcommand{\dW}{\mbox{W$^-\rightarrow$\lept\nubar}}
\newcommand{\dsele}{\mbox{\selem$\rightarrow$\chz e$^-$}}
\newcommand{\eeeell}{\mbox{\epair$\rightarrow$\epair\lpair}}
\newcommand{\eell}{\mbox{\epair\lpair}}
\newcommand{\llgam}{\mbox{$\ell\ell(\gamma)$}}
\newcommand{\nunugam}{\mbox{$\nu\bar{\nu}\gamma\gamma$}}
\newcommand{\acope}{\mbox{$\Delta\phi_{\mathrm{EE}}$}}
\newcommand{\nee}{\mbox{N$_{\mathrm{EE}}$}}
\newcommand{\eesum}{\mbox{$\Sigma_{\mathrm{EE}}$}}
\newcommand{\at}{\mbox{$a_{t}$}}
\newcommand{\spp}{\mbox{$p$/\Ebeam}}
\newcommand{\acoph}{\mbox{$\Delta\phi_{\mathrm{HCAL}}$}}

%-----------------------------------------
%  Variables for machine; math mode only
%-----------------------------------------
\newcommand{\roots}     {\sqrt{s}}
%----------------------------------------
%  Variables for events; math mode only
%----------------------------------------
%
%     Thrust
%
\newcommand{\thrust}    {T}
\newcommand{\nthrust}   {\hat{n}_{\mathrm{thrust}}}
\newcommand{\thethr}    {\theta_{\,\mathrm{thrust}}}
\newcommand{\phithr}    {\phi_{\mathrm{thrust}}}
\newcommand{\acosthr}   {|\cos\thethr|}
\newcommand{\thejet}    {\theta_{\,\mathrm{jet}}}
\newcommand{\acosjet}   {|\cos\thejet|}
\newcommand{\thmiss}    { \theta_{miss} }
\newcommand{\cosmiss}   {| \cos \thmiss |}

%-----------------------------------------
%     Energy, etc.
%-----------------------------------------
\newcommand{\Evis}      {E_{\mathrm{vis}}}
\newcommand{\Rvis}      {E_{\mathrm{vis}}\,/\roots}
\newcommand{\Mvis}      {M_{\mathrm{vis}}}
\newcommand{\Rbal}      {R_{\mathrm{bal}}}

%-----------------------------------------
% miscellaneous
%-----------------------------------------
\newcommand{\Ecm}{\mbox{$E_{\mathrm{cm}}$}}
\newcommand{\Ebeam}{\mbox{$E_{\mathrm{beam}}$}}
\newcommand{\ipb}{\mbox{pb$^{-1}$}}
\newcommand{\wrt}{with respect to}
\newcommand{\sm}{Standard Model}
\newcommand{\smb}{Standard Model background}
\newcommand{\smp}{Standard Model processes}
\newcommand{\smc}{Standard Model Monte Carlo}
\newcommand{\mc}{Monte Carlo}
\newcommand{\btb}{back-to-back}
\newcommand{\tp}{two-photon}
\newcommand{\tpb}{two-photon background}
\newcommand{\tpp}{two-photon processes}
\newcommand{\lp}{lepton pairs}
\newcommand{\vto}{\mbox{$\tau$ veto}}
\newcommand{\gsim}{\;\raisebox{-0.9ex}
           {$\textstyle\stackrel{\textstyle >}{\sim}$}\;}
\newcommand{\lsim}{\;\raisebox{-0.9ex}{$\textstyle\stackrel{\textstyle<}
           {\sim}$}\;}
\newcommand{\degree}    {^\circ}
\newcommand{\Rparity}{$R$-parity}
\newcommand{\Rp}  {$R_{p}$}  
\newcommand{\lb}  {$\lambda$}
\newcommand{\lbp} {$\lambda^{'}$}
\newcommand{\lbpp} {$\lambda^{''}$}

%-----------------------------------------
%     Acoplanarity
%-----------------------------------------
\newcommand{\phiacop}   {\phi_{\mathrm{acop}}}

%-----------------------------------------
%  Units
%-----------------------------------------

%-----------------------------------------
%  Bibliographic references
%-----------------------------------------
%
%     Journal names
%
\newcommand{\ZP}[3]    {Z. Phys. {\bf C#1} (#2) #3.}
\newcommand{\PL}[3]    {Phys. Lett. {\bf B#1} (#2) #3.}
\newcommand{\etal}     {{\it et al}.,\,\ }
\newcommand{\PhysLett}  {Phys.~Lett.}
\newcommand{\PRL} {Phys.~Rev.\ Lett.}
\newcommand{\PhysRep}   {Phys.~Rep.}
\newcommand{\PhysRev}   {Phys.~Rev.}
\newcommand{\NPhys}  {Nucl.~Phys.}
\newcommand{\NIM} {Nucl.~Instr.\ Meth.}
\newcommand{\CPC} {Comp.~Phys.\ Comm.}
\newcommand{\ZPhys}  {Z.~Phys.}
\newcommand{\IEEENS} {IEEE Trans.\ Nucl.~Sci.}
%
%     Collaboration names
%
\newcommand{\OPALColl}   {OPAL Collab.}
\newcommand{\DELPHIColl} {DELPHI Collab.}
\newcommand{\ALEPHColl}  {ALEPH Collab.}
\newcommand{\JADEColl}  {JADE Collab.}
%
%-------
%  etc
%-------
\newcommand{\onecol}[2] {\multicolumn{1}{#1}{#2}}
\newcommand{\ra}        {\rightarrow}   % \to can be used as well

\newcommand {\qqp}           {\mbox{$\mathrm{q\overline{q}^\prime}$}}
\newcommand {\qpq}           {\mbox{$\mathrm{q^\prime\overline{q}}$}}
\newcommand {\Z}             {\mbox{${\mathrm Z}^{0}$}}
\newcommand {\W}             {\mbox{W$^{\pm}$}}

%=======================================================================
%       Title Page
%=======================================================================

%-----------------------------------------------------------------------
\begin{titlepage}
%
%     Header
%
\begin{center}
    \large
    EUROPEAN LABORATORY FOR PARTICLE PHYSICS 
\end{center}
\begin{flushright}
    \large
    \PPEnum\\
%    \PNnum\\
    \Date
\end{flushright}

%
%     Main title
%
\begin{center}
    \huge\bf\boldmath
 Search for Stable and Long-Lived Massive Charged Particles  
%% with the OPAL Detector at LEP
 in $\ee$ Collisions at $\sqrt{s}=130 -183$~GeV
\end{center}
\bigskip
\bigskip
%
%     Author names
%
\begin{center}
\LARGE
The OPAL Collaboration \\
\bigskip
\bigskip
\bigskip
%%\large
%%{Authors: \Author \\
%%Editorial board: \EdBoard}
\end{center}
%
%     Abstract
%
\begin{abstract}%=======================================================
A search for stable and long-lived massive  particles of electric charge 
$|Q/e|$~=~1 or $2/3$, pair-produced in $\ee$ collisions at centre-of-mass energies
from 130 to 183~GeV, is reported by 
the OPAL collaboration at LEP. 
No evidence for production of these particles was observed
in a mass range between 45 and 89.5 GeV.
Model-independent 
upper limits on the production cross-section between
0.05 and 0.19~pb have been derived for scalar and spin-1/2 particles 
with charge $\pm$1.
Within the framework of the 
minimal supersymmetric model (MSSM),
this implies a lower limit of 82.5 (83.5) GeV on the mass of 
long-lived right- (left-)handed scalar muons and scalar taus. 
Long-lived charged leptons and 
charginos are excluded for masses below 89.5~GeV.
For particles with charge $\pm$2/3 the upper limits on the production
cross-section vary between 0.05 and 0.2~pb.
All limits, on masses and on cross-sections, are valid at the 95\% 
confidence level for particles with lifetimes
longer than 10$^{-6}$~s.
\end{abstract}%=========================================================
 
\bigskip
\smallskip
\begin{center}
{\large (Submitted to Physics Letters B)}
%% {\large \bf Draft \DraftVer, \DraftDate}
\end{center}
 
\smallskip
\begin{center}
%% {\large \bf Comments to \MailAddr}\\
%% {\large \bf by \TimeLimit}
\end{center}

\end{titlepage}
 
%=======================================================================
% OPAL author's list
%=======================================================================
%% \input{opal_authors}
\begin{center}{\Large        The OPAL Collaboration
}\end{center}\bigskip
\begin{center}{
%begin authorlist
K.\thinspace Ackerstaff$^{  8}$,
G.\thinspace Alexander$^{ 23}$,
J.\thinspace Allison$^{ 16}$,
N.\thinspace Altekamp$^{  5}$,
K.J.\thinspace Anderson$^{  9}$,
S.\thinspace Anderson$^{ 12}$,
S.\thinspace Arcelli$^{  2}$,
S.\thinspace Asai$^{ 24}$,
S.F.\thinspace Ashby$^{  1}$,
D.\thinspace Axen$^{ 29}$
% ...
%end authorlist
}\end{center}\bigskip
\bigskip
%begin institutes
$^{  1}$School of Physics and Astronomy, University of Birmingham,
Birmingham B15 2TT, UK
\newline
$^{  2}$Dipartimento di Fisica dell' Universit\`a di Bologna and INFN,
I-40126 Bologna, Italy
\newline
% ...
$^{  5}$Cavendish Laboratory, Cambridge CB3 0HE, UK
\newline
$^{  6}$Ottawa-Carleton Institute for Physics,
Department of Physics, Carleton University,
Ottawa, Ontario K1S 5B6, Canada
\newline
% ...
$^{  8}$CERN, European Organisation for Particle Physics,
CH-1211 Geneva 23, Switzerland
\newline
$^{  9}$Enrico Fermi Institute and Department of Physics,
University of Chicago, Chicago IL 60637, USA
\newline
% ...
$^{ 12}$Indiana University, Department of Physics,
Swain Hall West 117, Bloomington IN 47405, USA
\newline
% ...
$^{ 16}$Department of Physics, Schuster Laboratory, The University,
Manchester M13 9PL, UK
\newline
% ...
$^{ 23}$Department of Physics and Astronomy, Tel Aviv University,
Tel Aviv 69978, Israel
\newline
$^{ 24}$International Centre for Elementary Particle Physics and
Department of Physics, University of Tokyo, Tokyo 113, and
Kobe University, Kobe 657, Japan
\newline
% ...
$^{ 29}$University of British Columbia, Department of Physics,
Vancouver BC V6T 1Z1, Canada
\newline
% ...
\newline
%end notes

\newpage

%=======================================================================
\section{Introduction}
%=======================================================================
%% \input{dedx_intro}
%=======================================================================
%% \section{Introduction}
%=======================================================================
\label{sec:intro}

Most searches for new particles predicted by models
beyond the Standard Model (SM) assume that these particles decay
promptly at the primary interaction vertex due to their very short
lifetimes. These searches would not be sensitive to long-lived heavy particles
which do not decay in the detectors. 
However several models predict such long-lived particles.
For example, in the minimal supersymmetric model (MSSM)~\cite{ref:MSSM}, 
if the mass difference between 
the chargino and the lightest neutralino is smaller than a few hundred MeV,
the lightest  
chargino would have a lifetime sufficiently long to result in 
decays predominantly outside the detector.
In gauge-mediated supersymmetry, if the slepton is the
next-to-lightest supersymmetric particle, it could decay with a long lifetime
to a lepton plus a goldstino, if the SUSY-breaking energy scale is sufficiently
high~\cite{ref:gmssm}.
R-parity violating SUSY models~\cite{ref:rpv} also allow for long-lived heavy charged
particles. If the lightest supersymmetric particle is a slepton and the 
R-parity violating coupling is small ($\lambda
\leq 10^{-6}$), then the slepton decay length would be larger than a 
few meters, the typical size of tracking detectors at LEP.
If a fourth-generation heavy lepton ($m_{\rm L} > m_{\rm Z}/2$) 
exists~\cite{ref:heavy-lept}, the charged heavy
lepton would be stable if it is lighter than its neutral isodoublet 
partner.
%If a fourth-generation lepton family exists, and if the neutral heavy lepton is
%heavier than its charged counterpart, then the charged heavy lepton would be
%long-lived.
Some models beyond the SM would also predict the existence of
particles with fractional electric charge. As an example, 
leptoquarks~\cite{ref:lq} having a small coupling $\lambda_{\rm L, R}$
could be long-lived and possess fractional charge. Another example
could be long-lived hadronic states with fractional charge 
predicted by some modified QCD models~\cite{ref:dec-quarks}.

Previous searches for long-lived charged particles have been
performed by the LEP collaborations with data taken at the Z$^0$ 
resonance~\cite{ref:zo-stable}. 
The OPAL Limits on the production cross-section have been set at 
approximately 1.5 pb at 95\% CL for masses between
34 and 44~GeV. DELPHI and ALEPH~\cite{ref:lep2-stable} have also
analysed the data collected at centre-of-mass energies up to 172~GeV,
setting limits of 0.2-0.4 pb for masses between 45 and 86~GeV. 

This paper describes a search for long-lived particles 
X$^\pm$, with $m_{\rm X} > m_{\rm Z}/2$,
with charge $|Q/e|$~=~1 or 2/3, pair-produced in the
reaction $\ee \rightarrow {\rm X}^{+} {\rm X}^{-} (\gamma)$.
The data were collected by the OPAL detector 
during 1995-1997, at centre-of-mass energies 
of 130-136, 161, 172 and 183~GeV for a total integrated luminosity 
of 89.5~pb$^{-1}$.
Such particles distinguish themselves by their anomalous, high  or low
ionization energy loss, $\dedx$, in the tracking detector gas.
This search is therefore primarily based on the precise $\dedx$
measurement provided by the OPAL jet chamber. 
However, for particles of charge one, there is a large mass region 
where the measured $\dedx$ cannot distinguish them from 
ordinary particles. A complementary
search, based on the two-body kinematics,
is used to cover this mass region. 
No search was made for particles with $|Q/e|$~=~1/3 because their 
low ionization energy loss is too close to the jet chamber $\dedx$ 
measurement threshold.
The results obtained are valid for
particles with a lifetime longer than 10$^{-6}$~s.

%=======================================================================
\section{The OPAL Detector}
%=======================================================================
\label{sec:opaldet}

A complete description of the  OPAL detector can be found 
in Ref.~\cite{ref:OPAL-detector} and only a brief overview is given here.
The central detector consists of
a system of tracking chambers,
providing charged particle reconstruction
over 96\% of the full solid 
angle\footnote
   {The OPAL right-handed coordinate system is defined such that the $z$ axis is 
    in the
    direction of the electron beam, the $x$ axis is horizontal 
    and points towards the centre of the LEP ring, and  
    $\theta$ and $\phi$
    are the polar and azimuthal angles, defined relative to the
    $+z$- and $+x$-axes, respectively. The radial coordinate is denoted
    by $r$.}
inside a 0.435~T uniform magnetic field parallel to the beam axis. 
It consists of a two-layer
silicon microstrip vertex detector, a high-precision drift chamber,
a large-volume jet chamber and a set of $z$-chambers measuring 
the track coordinates along the beam direction. 

The jet chamber (CJ) is the most important detector for this analysis.
It is divided into 24 azimuthal sectors, each equipped with 159  
sense wires.  Up to 159 coordinate and $\dedx$
measurements per track are thus possible, with a 
precision of $\sigma_{r\phi} \approx 135 \, \mu$m and $\sigma_z \approx $~6~cm.
When a track is matched with $z$-chamber hits the uncertainty
on its $z_{0}$ coordinate is $\approx $~1~mm. 
%%% In a magnetic field of 0.435~T,
The jet and $z$-chambers, located inside the magnetic coil, 
provide a track momentum measurement with a resolution of
$\sigma_p/p \approx \sqrt{(0.02)^2 + (0.0015
\cdot p_t)^2}$ for tracks with the full number of hits 
($p_t$, in GeV, is the momentum transverse to the beam
direction) and a resolution on the ionization energy loss measurement 
of approximately $2.8\%$ for $\mumu$ events with a large number of
usable hits for $\dedx$~\cite{ref:dedx}.

A lead-glass electromagnetic 
calorimeter (ECAL) located outside the magnet coil
covers the full azimuthal range with excellent hermeticity
in the polar angle range of $|\cos \theta |<0.984$.
The magnet return yoke is instrumented for hadron calorimetry (HCAL)
covering the region $|\cos \theta |<0.99$ and is followed by
four layers of muon chambers.
Electromagnetic calorimeters close to the beam axis 
complete the geometrical acceptance down to 24 mrad
on both sides of the interaction point.
%except for the regions where a tungsten shield is present to protect
%the detectors from synchrotron radiation.
These include 
the forward detectors (FD), which are
lead-scintillator sandwich calorimeters and, at small angles,
the silicon tungsten calorimeters (SW)~\cite{ref:SW}.
The gap between the endcap ECAL and the FD
is instrumented with an additional lead-scintillator 
electromagnetic calorimeter,
called the gamma-catcher (GC).

\begin{figure}[htbp]
\centering
\epsfig{file=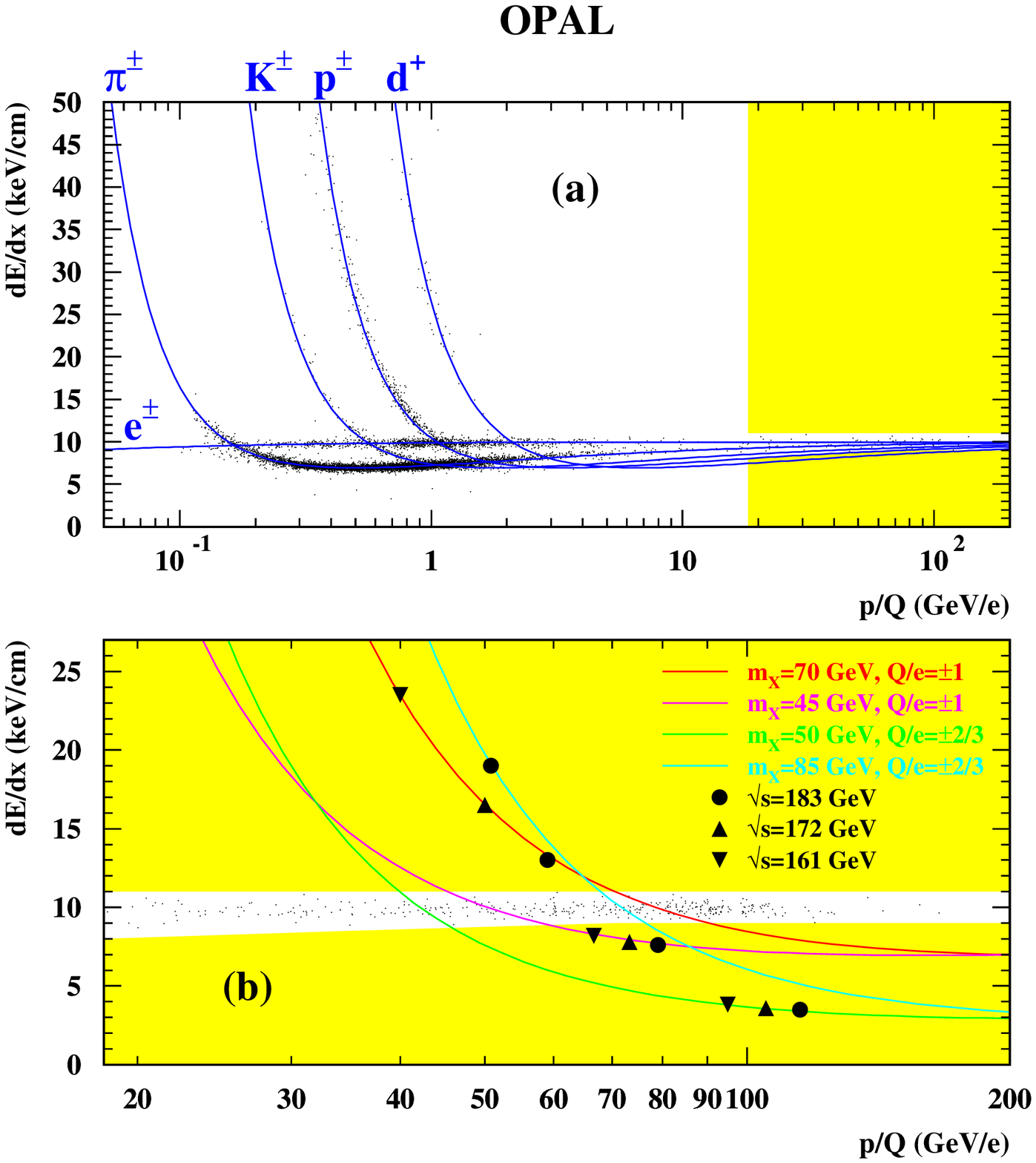,width=15cm}
\caption[]{\sl
  \protect{\parbox[t]{15cm}{
(a) The distribution of the ionization energy loss, $\dedx$, 
as a function of the apparent momentum $p/Q$
for data collected at $\sqrt{s}=183$~GeV.
The two hatched regions are the search regions at $\sqrt{s}=183$~GeV. 
The momentum lower limit is defined by the preselection cut
$p>0.1\sqrt{s}$.
A cutoff of $p_t>0.1$~GeV is made to reject low
momentum tracks curling in the jet chamber volume.\\
(b) Expanded view of the search regions. 
The theoretical curves for heavy long-lived particles are shown.
In $\ee\rightarrow\XX$ 
events, the momentum of the {\rm X}$^\pm$ particles of a given mass 
is fixed by $\sqrt{s}$. 
For $|Q/e|=2/3$, $m_{\rm X}=85$~GeV only the position at $\sqrt{s}=183$~GeV is visible,
while at
$\sqrt{s}=172$~GeV the $\dedx$ value lies outside the plot; $\sqrt{s}=161$~GeV
is below the pair-production threshold for this mass.
} } } 
\label{fig:dedx-demo}
\end{figure}

The ionization energy loss $\dedx$ produced by a charged particle 
is a function of $\beta\gamma=p/m$ and of the electric charge $Q$~\cite{ref:dedx}.
%%At low
%%values of $\beta\gamma$, $\dedx$ varies roughly as $1/\beta^2$, 
%%it reaches a minimum at
%%$\dedx\simeq 6.9$~keV/cm for $\beta\gamma\simeq 3.5$ and then rises up to the 
%%Fermi plateau at
%%$\dedx\simeq 9.9$~keV/cm for $\beta\gamma>10^3$. 
%% For $\ee$ and $\mumu$ events passing the preselection criteria, the mean 
%% $\dedx$ value is 
%% 9.85~keV/cm with an r.m.s. spread of 0.36~keV/cm.
In Figure~\ref{fig:dedx-demo}, the distribution of $\dedx$ as a function of 
the apparent momentum, $p/Q$, is shown. 
Standard particles of charge $\pm$1 (e, $\mu$, $\pi$, p, K) 
with high momentum ($p > 0.1 \sqrt{s}$ GeV) have $\dedx$ 
between 9 and 11~keV/cm. 
Massive particles with charge $\pm$1 are expected to yield
$\dedx>11$~keV/cm for high-mass values, $m_{\rm X} >0.36 \sqrt{s}$ 
(e.g., at 
$\sqrt{s}=183$~GeV, $m_{\rm X}=65$~GeV, one has $p=64$~GeV and $\dedx =11$~keV/cm), 
or $\dedx<9$~keV/cm for
low-mass values, $m_{\rm X} < 0.27 \sqrt{s}$ (e.g., at 
$\sqrt{s}=183$~GeV, $m_{\rm X}=50$~GeV, one has $p=77$~GeV and $\dedx =8$~keV/cm).
The $\dedx$ measurement therefore provides a good tool for particle 
identification in this high- and low-mass regions.
However, in the intermediate mass region, the $\dedx$ measurement does not
distinguish the signal from the background. In this region an 
analysis based on kinematic properties of pair-produced massive 
particles is used.

%=======================================================================
\section{Monte Carlo Simulation}
%=======================================================================
\label{sec:MC}

This section describes the Monte Carlo simulation of the signal and 
the background samples. 
All generated events have been processed
through the full simulation of the OPAL detector~\cite{ref:GOPAL};
the same event analysis chain has been applied to the simulated events
and to the data.

%=======================================================================
%% \subsection{Signal Monte Carlo Samples}
%=======================================================================

To generate the signal process $\ee \rightarrow {\rm X}^{+}{\rm 
X}^{-}(\gamma)$
different generators have been used.
Signal events of the type $\ee \rightarrow \sell^+
\sell^-(\gamma)$ ($\sell^{\pm}$ being a scalar lepton)
have been generated at four different energies 
($\sqrt{s}=133,161,172,183$~GeV) 
using SUSYGEN~\cite{ref:SUSYGEN}. The generated scalar leptons
are not allowed to decay, therefore simulating the signal from 
heavy charged stable scalar particles.
%% in events of the type $\ee\rightarrow \XX(\gamma)$.
Similarly, events of the type $\ee \rightarrow \LL(\gamma)$ and 
$\ee \rightarrow \QQ(\gamma)$, 
where L$^\pm$ are stable heavy leptons and Q,~$\bar{\mathrm Q}$ 
are stable heavy deconfined 
quarks with charge $\pm$2/3, have been generated at the same energies,
using the generator EXOTIC~\cite{ref:EXOTIC}.
All signal samples have been generated
with $m_{\rm X}$ ranging from 45~GeV to 90~GeV. 
Each sample contains 1000 events.
For the purpose of detector simulation and particle interactions,
$\sell^\pm$ and L$^\pm$ particles 
have been treated as heavy muons, while Q,~$\bar{\mathrm Q}$ 
as heavy stable hadrons with charge $\pm$2/3.
 
%=======================================================================
%% \subsection{Background Monte Carlo Samples}
%=======================================================================

The background has been estimated using simulations of all 
Standard Model processes (lepton-pair and multihadronic
processes, four-fermion processes,
two-photon processes). 
The Monte Carlo samples generated at $\sqrt{s}$~=~171 and 184~GeV are briefly
described below. 
Small differences in the centre-of-mass energies between data and the
simulated background Monte Carlo samples have a negligible effect on 
the analysis.
A detailed description of the Monte Carlo
samples generated at $\sqrt{s}$~=~133~GeV and
$\sqrt{s}$~=~161~GeV can be found in~\cite{ref:slept133,ref:slept161}. 
All background samples have an equivalent luminosity of at least ten times the data 
collected at each energy. 

The contribution to the background from two-fermion final states has been 
estimated using BHWIDE~\cite{ref:BHWIDE}
for the $\ee(\gamma)$ 
final states and KORALZ~\cite{ref:KORALZ} for the 
$\mumu(\gamma)$ and the $\tautau(\gamma)$ states. 
Multihadronic events, $\qq(\gamma)$,
have been simulated using PYTHIA~\cite{ref:JETSET1}. 

For the two-photon background, the PYTHIA~\cite{ref:JETSET1}, 
PHOJET~\cite{ref:PHOJET} and HERWIG~\cite{ref:herwig} Monte Carlo 
generators 
have been used for $\ee \qq$ final states
and the Vermaseren~\cite{ref:VERMASEREN} generator 
for all $\ee \ell^+ \ell^-$ final states.
All other four-fermion final states have been 
simulated 
with grc4f~\cite{ref:grace4f}, which takes into 
account interferences between all four-fermion diagrams.

%=======================================================================
\section{Data Analysis}
%=======================================================================
\label{sec:analysis}

Pair-produced stable or long-lived massive charged particles would manifest
themselves as events with two approximately back-to-back charged tracks. 
Particles of charge $\pm$1
are assumed to not interact strongly, hence
to not produce hadronic showers. Since they are massive, they 
do not produce electromagnetic showers either.
From these considerations, these events should be very similar to $\mumu$
events, the only difference being the higher mass of the particles.
%% The search for singly-charged particles is described in detail
%% in the following section.
%% from~\ref{sec:presel} to~\ref{sec:syst}.

No assumption is made for the interaction properties of long-lived 
particles with charge $\pm$2/3; hence no calorimetric signature is 
used in the search for such particles, which is 
described in subsection~\ref{sec:q23}.

%% \subsection{General preselection}
%% \label{sec:presel}

The events collected by OPAL at $\sqrt{s}=$130-183~GeV
have been required to pass a preselection which rejects events 
incompatible with the signal topology. 
%In the search for singly-charged particles, 
The following criteria are applied:
\begin{itemize}
\item[{\bf P1}]
{
Events are rejected if the total multiplicity of tracks in the central
detector and clusters in the ECAL is greater than 18.
%% High-multiplicity events are rejected.
Cosmic ray events are rejected~\cite{ref:leptpairs}, as well
as Bhabha scattering events~\cite{ref:robins}.
\item[{\bf P2}] 
{Events are required to contain exactly two tracks in the central 
detector satisfying basic
quality criteria\footnote
{The distance between the beam axis and the track at the point of
closest approach (PCA) must be less than 1~cm; the $z$-coordinate of the PCA 
must be less than 40~cm; the innermost hit of the track measured 
by the jet chamber 
must be closer than 75~cm to the beam axis.}
and having a momentum
$p>0.1\sqrt{s}$, a momentum transverse to the beam axis
$p_t>0.025\sqrt{s}$, polar
angle satisfying $|\cos{\theta}|<0.97$ and
at least 20 CJ hits usable for $\dedx$ measurement. 
The two selected tracks are 
required to have opposite electric charge.}
\item[{\bf P3}]
{To reduce background from two-photon interactions, 
the acoplanarity angle\footnote
   {The acoplanarity angle, $\phi_{\rm acop}$,
    is defined as 180$\degree$ minus 
    %% the complement of 
    the angle between the two tracks 
    in the $r-\phi$ plane.} between the two tracks is required to 
be $\phi_{\rm acop}<20\degree$ and
the total visible energy\footnote{The visible energy, 
the visible mass and the total transverse momentum of the event
are calculated using the method described in~\cite{ref:OPAL-Higgs}.} 
of the event
is required to be $E_{\rm vis}>0.2\sqrt{s}$. }
\item[{\bf P4}]
{To reduce background from events with initial state radiation, 
events containing an isolated ECAL cluster with an energy greater than
5~GeV are rejected. Isolation is defined as an angular separation  of
more than $15\degree$ from the closest charged track.}
\item[{\bf P5}]
{It is required that 
$\frac{E_1}{p_1}+\frac{E_2}{p_2}<0.2$, where $E_{1,2}$ denotes the 
energies
of the ECAL clusters associated to the two selected tracks, to further reduce
the contribution from Bhabha scattering events. 
Moreover, in a cone of $10\degree$ half-opening angle around each of 
the two selected tracks, no other
tracks with $p>0.5$~GeV and no unassociated clusters with $E>3$~GeV
should be found.}
%%\\
%%These requirements will not be applied in the search for fractionally-charged
%%particles.
}
\end{itemize}

After these preselection criteria, the background is dominated by 
$\ee \rightarrow \mumu$ 
events, with a
small contribution from $\ee \rightarrow \tautau$ and two-photon $\ee\mumu$ events.
The effect of these preselection cuts on the samples at $\sqrt{s}=183$~GeV is shown 
in Table~\ref{tab:cutflow183}. 

%% The high mass of the particles in the signal is then exploited to 
%% distinguish signal from background. 

%%%%%%%%%%%%%
\subsection{Search for particles with charge $\pm$1}
\label{sec:dedx}

Two complementary methods are 
adopted, depending on the mass $m_{\rm X}$ of the signal particle.
In the high-mass ($m_{\rm X} >0.36 \sqrt{s}$) and
low-mass ($m_{\rm X} <0.27 \sqrt{s}$)
regions, the $\dedx$ measurement is 
used to distinguish the signal from the background.
The preselected events are retained if they satisfy 
the following requirements on $\dedx$:
\begin{itemize}
\item[{\bf A1}]
{Both high-momentum tracks must have either $\dedx>11$ keV/cm or 
$\dedx<9$ keV/cm. }
\item[{\bf A2}]
{The probability that either of the two $\dedx$ measurements originates from one 
of the standard particles (e, $\mu$, $\pi$, p, K) must be less than 10\%.}
%% The $\dedx$ simulation in the Monte Carlo samples has been checked
%% against Bhabha scattering and di-muon data events.
\end{itemize}

After this selection ({\bf A1}-{\bf A2}), no candidate survives in any of the data sets.
The total background is estimated to be 0.03 events at $\sqrt{s}=130-136$~GeV
and less than $10^{-2}$ events at any other energy
(see Tables~\ref{tab:cutflow183}, \ref{tab:cands}).

%% \subsection{Intermediate mass event selection for singly-charged particles}
%% \label{sec:kine}

A complementary analysis is used for 
masses in the intermediate mass range ($0.27<m_{\rm X}/\sqrt{s}<0.36$), where the 
$\dedx$ measurement does not 
provide adequate separation between the signal and the di-lepton background. 
% The overall selection will be the combined result of the two methods.
In events of the type $\ee \rightarrow \XX$, since 
$m_{\rm X}>m_{\rm Z}/2$ is assumed,
initial state radiation is suppressed. Hence, each particle should have an energy 
close to the beam energy and $m_{\rm X}$ can be 
estimated for each track as $\widehat{m}_{\rm X}=\sqrt{s/4-p^2}$. 
However, for SM
events with two tracks and missing energy (e.g., events with neutrinos, with
initial state radiation or two-photon interactions), a large overestimate 
of $\widehat{m}_{\rm X}$ could occur, providing a high background in the accepted
mass range. 
The following selection criteria are designed to reject these backgrounds.

\begin{figure}[htbp]
\centering
\epsfig{file=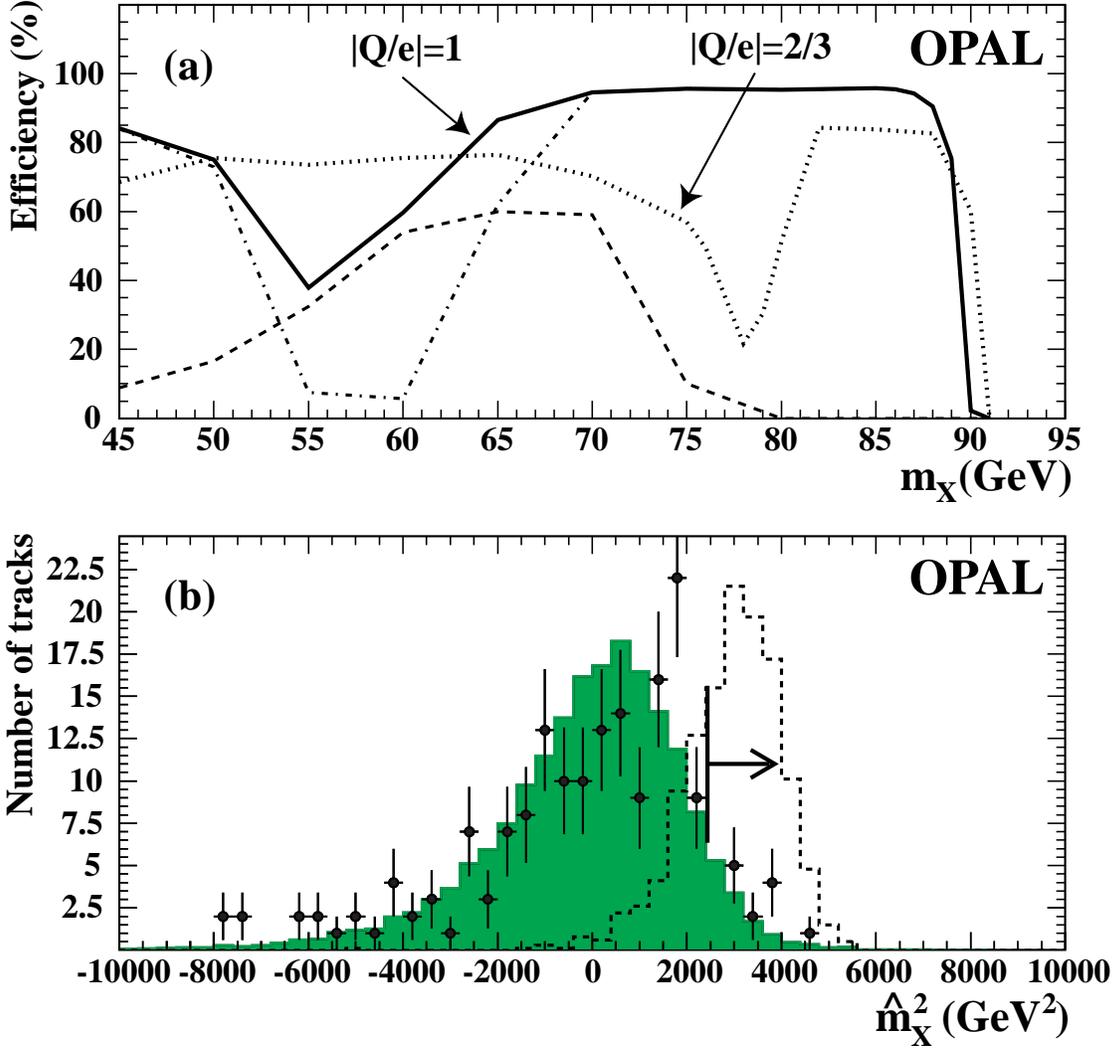,width=15cm}
\caption[]{\sl
  \protect{\parbox[t]{15cm}{
(a) Selection efficiency as a function of the mass $m_{\rm X}$ for
$|Q/e|$~=~1 and 2/3 particles, at $\sqrt{s}=183$~GeV. 
The dashed-dotted and dashed lines represent the efficiency
of the $\dedx$-based and kinematic selections, respectively, 
in the case of the spin-0 particles of charge $\pm$1.
The solid line 
refers to the combined selection.
The dotted line represents the overall efficiency of the search for
spin-1/2 particles of charge $\pm$2/3.\\
(b) The distribution of the $\widehat{m}_{\rm X}^2$, after cut
{\bf B4}, for data collected at $\sqrt{s}=183$~GeV (points with error 
bars), for all 
simulated backgrounds (grey histogram)
% (grey~=~$\mumu$; black~=~$\tautau$) 
and for a simulated signal (dashed line) with $m_{\rm X}$=55~GeV.
The signal histogram is arbitrarily normalised.
The cut {\bf B5} is indicated by the arrow pointing to the accepted region.
No event is selected when both selected tracks are required to pass this cut.
} } } 
\label{fig:cuts}
\end{figure}

\begin{itemize}
\item[{\bf B1}]
{To reject background events with two tracks and neutrinos, 
the acoplanarity angle of the two tracks is required to be 
$\phi_{\rm acop}<1\degree$.}
\end{itemize}
\begin{itemize}
\item[{\bf B2}]
{To reduce the background from two-photon interactions and radiative events,
the total visible energy of the event is required to be 
$E_{\rm vis}>0.6\sqrt{s}$ and the total momentum along the beam direction
is required to be $|p_z^{tot}|<0.2 \sqrt{s}$.}
\item[{\bf B3}]
{The polar angle
of the missing momentum vector must satisfy $|\cos \theta_{miss}| < 0.8$.
The total energies measured in the forward detectors (FD, SW, GC) 
must satisfy: $E_{\rm FD}<5$~GeV, $E_{\rm SW}<5$~GeV, $E_{\rm GC}<5$~GeV.
The total energy deposited in the HCAL is required to be less than 15~GeV.
The last two requirements introduce relative inefficiencies of at most 3.2\% and
less than 1\%, respectively, due to
electronic noise in these detectors; the final efficiencies are corrected
by these factors.}
\end{itemize}
%
%If a track is too close to a jet chamber anode plane, its
%momentum measurement is affected by a large systematic uncertainty which is
%not well modelled in the detector simulation. For this reason:
\begin{itemize}
\item[{\bf B4}]
{Both selected tracks are required to be at least $0.5\degree$ away from the CJ 
anode planes in the $r-\phi$ projection,
to avoid momentum mismeasurements.}
\end{itemize}
At this stage the background is mainly due to $\mumu$ events with a smaller
contribution from $\tautau$ events where both $\tau$'s decay into muons. 
The intermediate-mass signal is
then selected by requiring that:
\begin{itemize}
\item[{\bf B5}]{For both tracks, $\widehat{m}_{\rm X}^2 > (0.27\sqrt{s})^2$. 
At $\sqrt{s}$=130-136, 161 GeV this cut is not stringent enough, therefore
it is required that
$\widehat{m}_{\rm X}^2 > (45~{\mathrm GeV})^2$.
%% Since this selection would yield a high background at
%% $\sqrt{s}\leq 161$~GeV,  
%% is required for these energies.
Figure~\ref{fig:cuts}(b) shows the distribution of 
$\widehat{m}_{\rm X}^2$ for both selected tracks, after cut {\bf B4}
at $\sqrt{s} = 183$~GeV.}
\end{itemize}
The effect of cuts {\bf B1-B5} for data and simulated events
can be seen in Table~\ref{tab:cutflow183}.

An event is selected as a candidate if it satisfies either of the two 
sets of selection criteria ({\bf A1}-{\bf A2}) or ({\bf B1}-{\bf B5}), following the 
preselections ({\bf P1}-{\bf P5}). 
The overall detection efficiency for spin-0 particles 
is shown in Figure~\ref{fig:cuts}(a),
as a function of $m_{\rm X}$, for $\sqrt{s}=183$~GeV. 
It exceeds 37\% even in the mass interval where the $\dedx$-based 
selection is inefficient.
At lower centre-of-mass energies the efficiencies have similar values
and behaviours.
%more precisely, they can be 
%considered simply a function of $m_X/\sqrt{s}$ with an uncertainty 
%of less than 5\%.
For spin-1/2 particles, the efficiencies are 2-9\% lower due to the 
different angular distribution of the tracks.

%After this selection one candidate survives in the $\sqrt{s}=130$~GeV data and
%one in the $\sqrt{s}=161$~GeV data; no candidates in the other data samples.
%The expected background rates at the energy points of 133-136, 161, 172,
%183~GeV are respectively 0.46, 0.22, 0.31 and 0.97 events

After this selection one candidate survives in the $\sqrt{s}=161$~GeV data set
while no candidates are found in the other data samples.
The expected backgrounds at 130-136, 161, 172,
183~GeV energies are 0.46, 0.22, 0.31 and 1.01 events, respectively (see Table~\ref{tab:cands}).

% The candidate at $\sqrt{s}=130$~GeV have tracks' momenta $p_1=37.2\pm 14.6$~GeV
% and $p_2=46.9\pm 6.0$~GeV, leading to estimated masses of 53.3~GeV and 45.0~GeV
% respectively --- the second one being just 4~MeV above the lower threshold of
% 45~GeV. The difference in the two momenta, together with an acollinearity of 
% the two tracks of xx~degrees, suggests that this event is affected by missing 
% energy. 
The candidate at $\sqrt{s}=161$~GeV has selected tracks with momenta 
$p_1=(51.8\pm 3.8$)~GeV
and $p_2=(46.7\pm 2.9$)~GeV, leading to estimated masses of ($61.6\pm 3.2$)~GeV and 
($65.6\pm 2.1$)~GeV, respectively.
A possible interpretation for this event in the SM is 
a $\tautau$ event where both $\tau$'s decay into $\mu$'s.

\begin{table}[hbt]
\centering
\begin{tabular}{|l||r||r||r|r|r|r|r||r|r|r|}
\hline
\multicolumn{1}{|c||}{Cuts}      & \multicolumn{1}{c||}{Data}   &
\multicolumn{6}{c||}{Background Simulation} & \multicolumn{3}{c|}{Signal MC (\%)}\\

\cline{3-11}
%------------------------------------------------------------------
           &
& \multicolumn{1}{c||}{Total}
& \multicolumn{1}{c|}{$\ee$}
& \multicolumn{1}{c|}{$\mumu$}
& \multicolumn{1}{c|}{$\tautau$}
& \multicolumn{1}{c|}{$\ee\ellell$}
& \multicolumn{1}{c||}{Others}
& $\epsilon_{45}$ & $\epsilon_{55}$ & $\epsilon_{80}$ \\
\hline%-----------------------------------------------------------------
 %              data   bkg.tot   ee->ee    ee->mm    ee->tt    gg->ll    others eff1 eff2 eff3
 {\bf P1--2}&  1595&  1517.04&   806.97&   378.71&    90.58&   147.05&    93.73&97.0&97.5&97.7\\
 {\bf P3--4}&  1186&  1158.97&   636.45&   309.29&    66.48&   109.11&    37.63&92.7&93.1&95.8\\
 {\bf P5   }&   369&   357.11&     0.00&   288.82&     5.07&    56.53&     6.70&92.5&92.8&95.4\\
\hline%-----------------------------------------------------------------
 {\bf A1,A2}&     0&     0.00&     0.00&     0.00&     0.00&     0.00&     0.00&84.0& 7.5&95.4\\
\hline%-----------------------------------------------------------------
 {\bf B1   }&   297&   284.40&     0.00&   247.73&     2.55&    32.30&     1.81&88.8&90.2&93.0\\
 {\bf B2--4}&    88&    92.61&     0.00&    92.30&     0.25&     0.00&     0.06&61.1&61.4& 0.2\\
 {\bf B5   }&     0&     1.01&     0.00&     0.85&     0.15&     0.00&     0.01& 8.7&32.0& 0.0\\
\hline%-----------------------------------------------------------------
\hline%-----------------------------------------------------------------
 {\bf Comb.}&     0&     1.01&     0.00&     0.85&     0.15&     0.00&     0.01&84.1&37.5&95.4\\
\hline%-----------------------------------------------------------------
\end{tabular}
\caption[]{\sl
  \protect{\parbox[t]{15cm}{
%This table refers to the analysis at $\sqrt{s}=183$~GeV.
The numbers of events remaining after each cut 
for data collected at $\sqrt{s}=183$~GeV and 
for various Monte Carlo
background processes normalised to the
integrated luminosity of the data 
(``{\rm Others}'' refers to $\/ \ee \rightarrow \qq$, 
and $\/ \ee \rightarrow$ four-fermion processes).
%and the total background rate is compared with data 
%(second column) after each cut.
In the last three columns the efficiencies for $\sell^{\pm}$ 
are given (in percent)
for $m_{\rm X}=45, 55, 80$~GeV.
The last line gives the results after combining the two analyses
described in the text.
The relative rate of data and backgrounds and the composition of the background
at lower energies is similar to this case.
} } }
\label{tab:cutflow183}

\end{table}

\subsection{Systematic uncertainties}
\label{sec:syst}

The main systematic errors affecting the signal detection efficiencies 
%% (independently for each data set)
are listed below and their values are summarised in Table~\ref{tab:syst}.
These are: the statistical error from the 
Monte Carlo samples,
%%$\sell^+\sell^-$, $\LL$ and $\QQ$ Monte Carlo simulation is 0.9--1.7\%;
the uncertainty due to the linear interpolation of the efficiencies,
%%is estimated to be 2\%.
the evaluation of the integrated luminosity~\cite{ref:PLB391},
%%0.5--0.9\%, determined by using the SiW luminometer~\cite{ref:PLB391};
the measurement of $\phi_{\rm acop}$ and
the measurement of $E_{\rm vis}$ and $p_z^{\rm tot}$.
The uncertainty due to the $\dedx$ measurement (cuts {\bf A1} and {\bf A2})
is estimated as follows: for each Monte Carlo signal event, 
if both tracks have $\dedx>11$~keV/cm ($\dedx<9$~keV/cm),
both $\dedx$ values are decreased (increased) by their errors, then 
the selection cuts are applied to the modified event. The resulting efficiency 
is therefore lower; the variation in the efficiency is taken as the systematic 
error.
The uncertainty due to the momentum measurement (cut {\bf B5}) is estimated in
the same way, but instead of varying the $\dedx$ values, both track momenta are
increased by their errors. Again, the resulting decrease in efficiency is taken
as the systematic error. 
The uncertainty due to cut {\bf B5}
is estimated to be at most 30\% in the intermediate mass region, where it 
dominates all the systematic effects.

%% Since no Monte Carlo signal samples have been generated 
%% for $\ee\rightarrow\LL$ and $\ee\rightarrow\QQ$ events 
%% at $\sqrt{s}\leq 161$~GeV, 
%% the efficiencies for such processes have been rescaled from those calculated 
%% at higher energies from Monte Carlo, assuming that they depend only on 
%% $m_{\rm X}/\sqrt{s}$. 
%% The estimated uncertainty arising from this procedure is 
%% evaluated to be at most 5\%.

The systematic errors are assumed to be independent, and
the total systematic error is calculated as the quadratic sum
of the individual errors. 

The uncertainty on the background arises essentially from the cuts {\bf A1, A2} 
and {\bf B5} and is studied in the same way as described above. At 
$\sqrt{s}$~=~183~GeV the uncertainty on the background is $\pm$ 0.69 
events.
%% for the systematic 
%% uncertainty on the selection efficiency. 
In computing the results, no 
background subtraction is performed.

%Conservatively, the
%uncertainty providing the smallest value of the background was chosen to
%perform background subtraction in computing limits, following 
%Reference~\cite{ref:cousins}.

\begin{table}[hbt]
\centering
\begin{tabular}{|c|c|c|}
\hline%------------------------------------------------------
Quantity & \multicolumn{2}{c|} {Systematic uncertainty (\%)} \\
\cline{2-3}
      & ($m_{\rm X}/\sqrt{s}<0.27$ or $m_{\rm X}/\sqrt{s}>0.36$) &
                  $\;\;\;\;\;\;\;$ ($0.27 < m_{\rm X}/\sqrt{s} <0.36$) $\;\;\;\;\;\;\;$ \\
\hline%------------------------------------------------------
MC statistics   &  \multicolumn{2}{c|} {0.9-1.6} \\
Luminosity      & \multicolumn{2}{c|} {0.5-0.9} \\
$\phi_{\rm acop}$   &  \multicolumn{2}{c|} {0.1}    \\
$E_{\rm vis}$, $p_z^{tot}$ &  \multicolumn{2}{c|} {0.5-0.9} \\
\cline{2-3}
Interpolation   &  2.0     & 5.0    \\
$\dedx$         &  0.0-1.8 & 1.8-8.9 \\
$\widehat{\rm m}_{\rm X}$  &  0.0-1.7 & 1.7-30.0 \\
%% Efficiency rescaling &  5.0      \\
\hline%------------------------------------
 Total          & 2.1-3.5 &  5.7-30.0 \\
\hline%------------------------------------
\end{tabular}
\caption[]{\sl
  \protect{\parbox[t]{15cm}{
Relative systematic uncertainties associated with the various quantities 
used. 
The systematic errors vary slightly with centre-of-mass energy, but
strongly with $m_{\rm X}$. The uncertainty in the $\widehat{m}_{\rm X}$
determination dominates in the intermediate mass region but has a small effect
in the mass regions where the $\dedx$-based selection is efficient.
%% The last line contains the total systematic error.

} } }
\label{tab:syst}
\end{table}

\subsection{Search for particles with charge $\pm 2/3$}
\label{sec:q23}

The search for stable and long-lived massive particles 
with charge $\pm$2/3 has been performed 
%over the full data sample at $\sqrt{s}=130-183$~GeV. 
without assumptions for the interaction properties of such particles. 
Therefore, requirements 
on energy deposits in the calorimeters are not applied. 
The selection is based on $\dedx$ information only.
The applied cuts are {\bf P1} through {\bf P4}, followed by {\bf A1} and {\bf A2}.

The ionization energy loss $\dedx$ scales roughly as $(Q/e)^2$; hence particles
having $|Q/e| =$~2/3 would generally have $\dedx$ values smaller than those
of charge $\pm$1 particle tracks.
For masses $m_{\rm X}<0.41\sqrt{s}$ (e.g., at 
$\sqrt{s}=183$~GeV, $m_{\rm X}=75$~GeV, one has $p=79$~GeV and $\dedx =8$~keV/cm) 
or $m_{\rm X}>0.44\sqrt{s}$ (e.g., at 
$\sqrt{s}=183$~GeV, $m_{\rm X}=82$~GeV, one has $p=61$~GeV and $\dedx =13$~keV/cm),
the $\dedx$ 
for the two tracks of the signal events would be well separated from that of 
SM di-lepton
events, as can be seen from Figure~\ref{fig:dedx-demo}.
% in these two
% regions the selection efficiency is higher than 60\%.
The narrow intermediate region
%, where the $\dedx$ of the signals
% overlaps that of normal particles, 
is not covered by any kinematical analysis.
%% as that used for singly-charged particles. 
However, the efficiency of selections {\bf A1}-{\bf A2} 
remains larger than 20\%, because the small variation
of track momenta due to initial state radiation can yield a 
variation of $\dedx$ which is larger than the excluded range.

After this selection, one candidate remains in the data sample at
$\sqrt{s}=183$~GeV, and 
no candidates are left in any of the other data sets.
The background at $\sqrt{s}=183$~GeV is estimated to be less than 0.3 events;
the total background, summed over all energies, is estimated 
to be less than 0.45 events, as shown in Table~\ref{tab:cands}.  
The two tracks of the candidate event have $\dedx$ values of $11.06\pm 0.35$~keV/cm and 
$11.06\pm 0.34$~keV/cm, and momenta ($70.9\pm 7.0$)~GeV
and ($64.7\pm 6.6$)~GeV, yielding estimated masses of
$(93.4 \pm 9.6)$~GeV and $(88.0 \pm 9.8)$~GeV, respectively
(the particle mass is evaluated from the momentum and the
estimate of $\beta\gamma$ obtained from the $\dedx$ measurement,
assuming $|Q/e|$=2/3).
However in this event the mismatch in polar angle
between the two selected tracks and the ECAL clusters gives evidence of 
% a Bhabha event in which the tracks have been
% reconstructed with wrong polar angles, 
a reconstruction error of both tracks. 
This effect
is described by the OPAL detector simulation program and would
explain an overestimate of the $\dedx$ values by affecting the 
path length.

The systematic uncertainties on the selection efficiencies
are estimated as described in the previous 
section and they are similar in size.

%=======================================================================
\section{Results}
%=======================================================================
\label{sec:results}
The numbers of candidates found in the search for particles 
with charge $|Q/e|$=1 and $|Q/e|$=2/3 are summarised in 
Table~\ref{tab:cands}, together
with the expected backgrounds. 
The data show no excess
above the expected background from Standard Model processes. 
Therefore model-independent cross-section upper limits 
have been computed for 
the pair-production
of massive charged long-lived particles.

\begin{table}[hbt]
\centering
\begin{tabular}{|c|c||cc|c|cc|c||c|c|}
\hline%------------------------------------
           &         & \multicolumn{6}{c||}{$|Q/e|=1$ search}           
                     & \multicolumn{2}{c|}{$|Q/e|=2/3$ search} \\
\cline{3-10}
$\sqrt{s}$ & Lumi.   & \multicolumn{3}{c|}{candidates} & \multicolumn{3}{c||}{background} & cand. & back. \\
(GeV)    &(pb$^{-1}$)& $\dedx$ & kin. & comb.          & $\dedx$ & kin. & comb.           &       &       \\
\hline%------------------------------------
130-136    & 11.0    & 0 & 0 & 0 &    0.03  & 0.42 & 0.46 & 0 & $<$0.08 \\
  161      & 10.1    & 0 & 1 & 1 & $<$0.005 & 0.22 & 0.22 & 0 & $<$0.02 \\
  172      & 10.3    & 0 & 0 & 0 & $<$0.005 & 0.31 & 0.31 & 0 & $<$0.07 \\
  183      & 58.1    & 0 & 0 & 0 & $<$0.005 & 1.01 & 1.01 & 1 & $<$0.27 \\
\hline%------------------------------------
 Total     & 89.5    & 0 & 1 & 1 &    0.03  & 1.96 & 2.00 & 1 & $<$0.44 \\
\hline%------------------------------------
\end{tabular}
\caption[]{\sl
  \protect{\parbox[t]{15cm}{
The number of candidate events and the expected background
at all energies, for the search for $|Q/e|=$~1 and 2/3 
particles. For $|Q/e|=1$, the result of both the $\dedx$-based and the kinematic
selections are shown separately and then combined.
In the second column, the integrated luminosity is given for each energy.
} } }
\label{tab:cands}
\end{table}

All data collected so far at energies above the Z$^0$ peak 
($\sqrt{s} = 130-183$~GeV) have been combined, 
assuming s-channel production and
therefore an energy dependence of the cross-section of
$\beta^3/s$ for spin-0 particles and $\frac{\beta}{s}(1-\frac{\beta^2}{3})$ 
for spin-1/2 particles, where $\beta=p/E\simeq\sqrt{1-4m_{\rm X}^2/s}$.

For the production of particles of charge $\pm$1, the candidate event at
$\sqrt{s}=161$~GeV has been taken into account, assuming a mass of 
($64.4\pm 1.8$)~GeV, equal to the weighted average of the two measured masses,
mentioned in Section~\ref{sec:dedx}.
Similarly, for particles of charge $\pm$2/3, the candidate event at
$\sqrt{s}=183$~GeV is included with a mass of 
($90.7\pm 6.9$)~GeV.
In evaluating upper limits, the candidates are counted in mass intervals 
centred on their central values and $\pm 2\sigma$ wide.
The total systematic error is incorporated into the limits, 
%% and background subtraction is performed, 
following the prescription of Ref.~\cite{ref:cousins}.
No background subtraction is performed.

\begin{figure}[htbp]
\centering
\epsfig{file=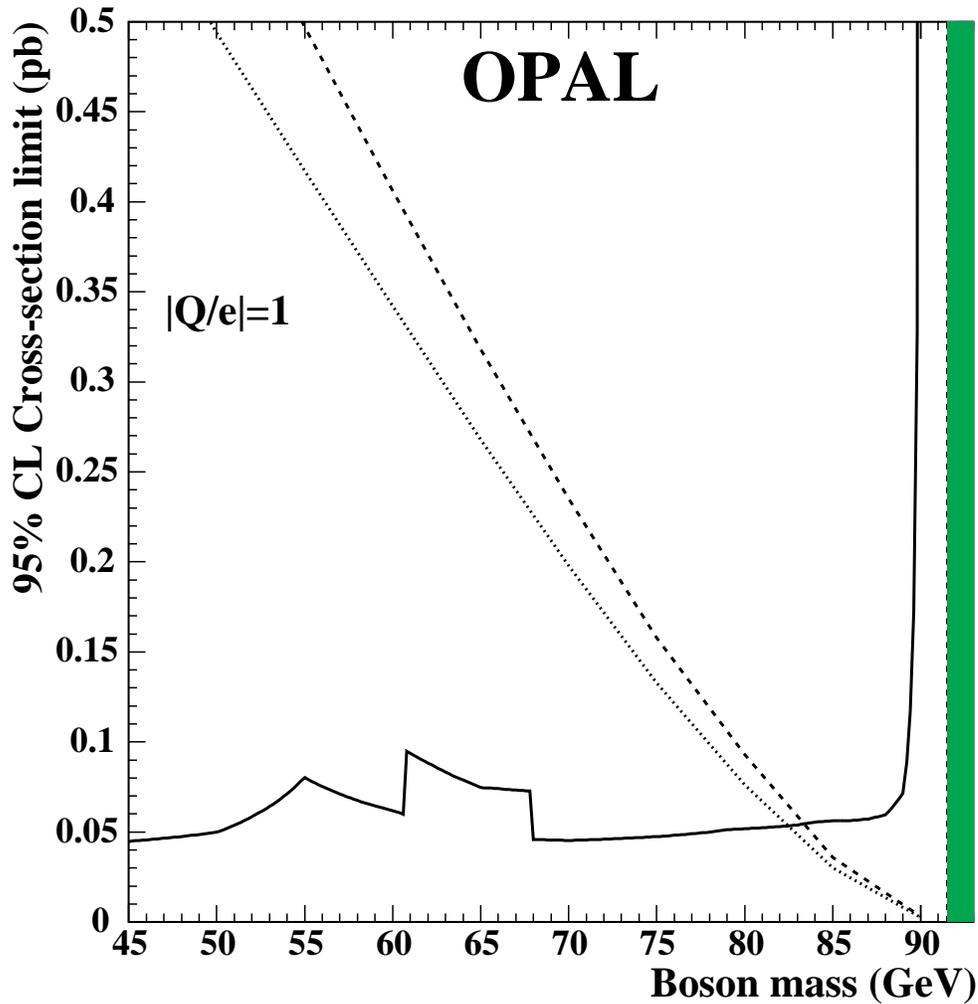,width=15cm}
\caption[]{\sl
  \protect{\parbox[t]{15cm}{
Model-independent 95\% CL upper limits on the pair-production
cross-section of spin-0 heavy long-lived particles of charge $\pm$1 as a function
of their mass (solid line). In calculating the upper limit, the candidate
is considered as described in the text.
The MSSM predicted cross-sections
for right-handed (dotted line) and left-handed (dashed line) smuons 
and staus are also shown. 
The 95\% CL lower limits on the
masses of these sleptons are at the crossing point between the experimental
and theoretical curves.
The grey region 
is kinematically inaccessible.
} } } 
\label{fig:xsect-scalar}
\end{figure}

In Figure \ref{fig:xsect-scalar}, the 95\% CL upper limit 
on the cross-section at $\sqrt{s} = 183$~GeV is shown for spin-0 particles. 
The 95\% CL upper limit on the pair-production cross-section varies 
from 0.05 to 0.1 pb in the mass range $45 < m_{\rm X} < 89.5$~GeV.
The cross-section limits are compared with the predicted 
cross-sections
for pair-production of right- and left-handed smuons and staus. For these two
slepton species, the production cross-section does not depend on the MSSM
parameters but only on the slepton mass.
The 95\% CL lower limits on the mass of right- and left-handed smuons and staus of
82.5~GeV and 83.5~GeV, respectively, are derived,
as shown in Figure~\ref{fig:xsect-scalar}. 

\begin{figure}[htbp]
\centering
\begin{tabular}{cc}
\epsfig{file=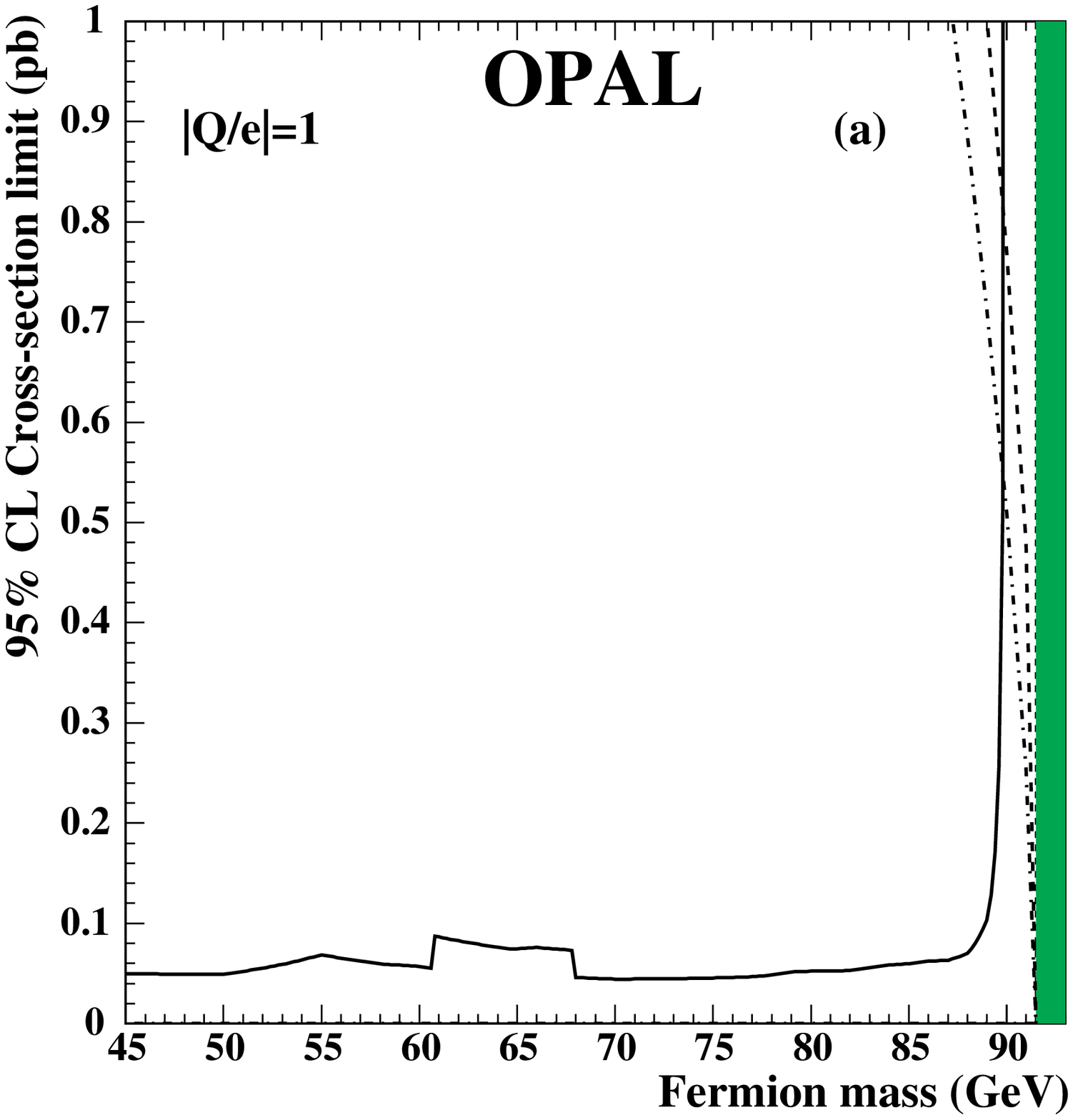,width=8.5cm} &
\epsfig{file=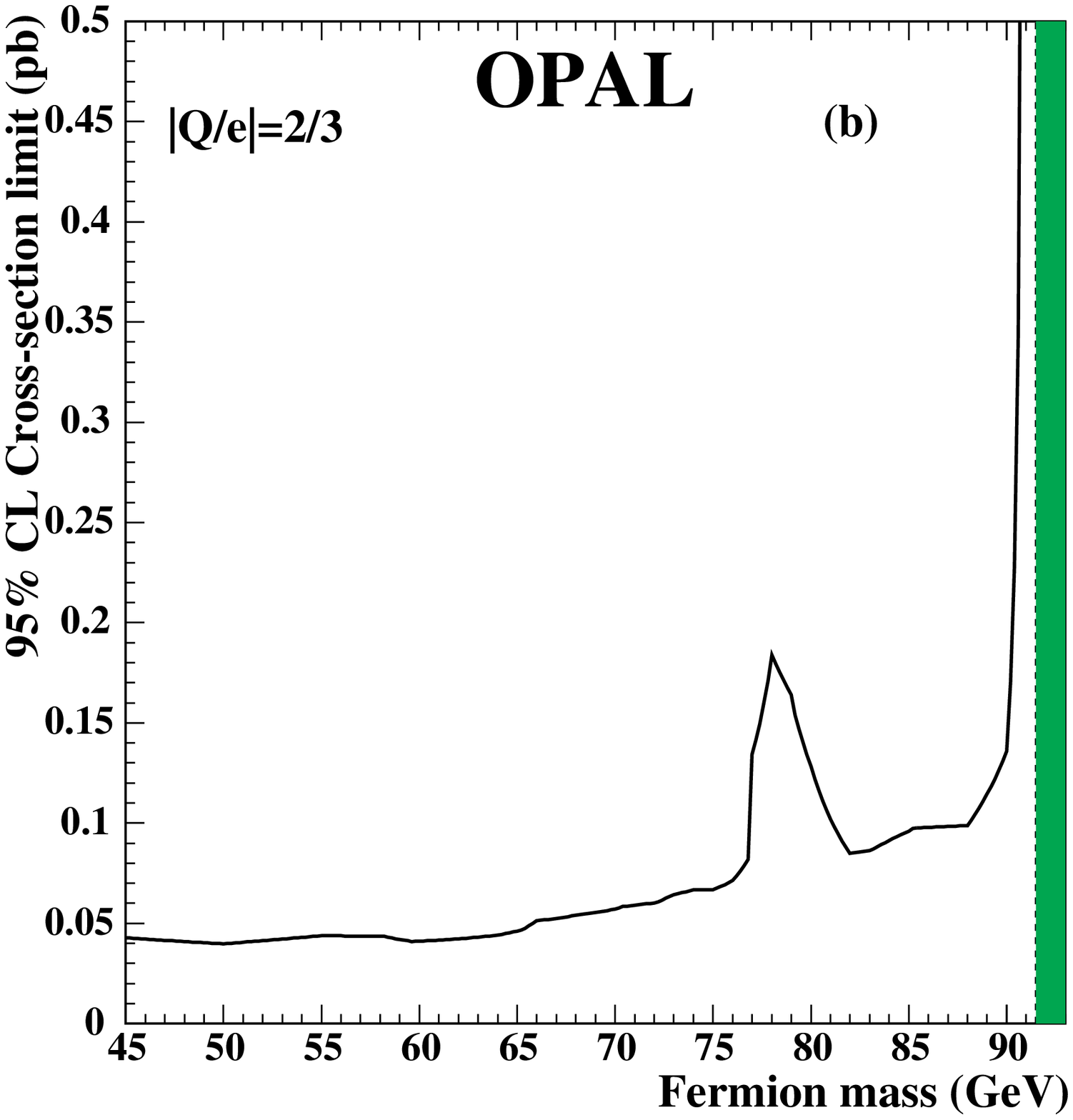,width=8.5cm}
\end{tabular}
\caption[]{\sl
  \protect{\parbox[t]{15cm}{
(a) Model-independent 95\% CL upper limits on the pair-production
cross-section of spin-1/2 heavy long-lived particles of charge $\pm$1 as a 
function of their mass (solid line). 
In calculating the upper limit, the candidate
is considered as described in the text.
The MSSM predicted cross-sections
for charginos (dashed line) and heavy leptons (dashed-dotted line)
are also shown.
The 95\% CL lower limits on the
masses of these particles are at the crossing point between the experimental
and theoretical curves.
(b) Model-independent 95\% CL upper limits on the pair-production
cross-section of spin-1/2 heavy particles of charge $\pm$2/3 as a function
of their mass (solid line). 
In both plots the grey region
is kinematically inaccessible.

} } } 
\label{fig:xsect-fermion}
\end{figure}

Figure \ref{fig:xsect-fermion}(a) shows the 95\%~CL 
upper limit on the cross-section at $\sqrt{s} = 183$~GeV for spin-1/2
particles; the limit varies 
from 0.05 to 0.19 pb in the mass range $45 < m_{\rm X} < 89.5$~GeV.
%, obtained combining data at $\sqrt{s}=172,~183$~GeV only.
This limit is compared
with the predicted cross-sections for chargino production
and heavy charged lepton production.
The MSSM parameters have been chosen to minimise 
the predicted chargino cross-section at every chargino
mass value (assuming a heavy sneutrino, $m_{\sneutrino} > 500$~GeV),
without any restriction on the mass of the lightest neutralino.
Therefore a 95\%~CL lower limit on
the masses of long-lived charginos is derived at 89.5~GeV for every
choice of the MSSM parameters.
The 95\%~CL lower limit on the heavy charged lepton mass is 89.5~GeV.

Figure \ref{fig:xsect-fermion}(b) shows the 95\%~CL upper limit on
the cross-section at $\sqrt{s} = 183$~GeV for spin-1/2 particles 
of charge $\pm$2/3; the limit varies between 0.05 to 0.2 pb in 
the mass range $45 < m_{\rm X} < 89.5$~GeV.
For spin-0 leptoquarks the 
cross-section upper limits would be slightly better than those 
calculated from spin-1/2 particles.
These limits can be compared with leptoquarks and deconfined quarks models 
which provide cross-section predictions. 
%,
%obtained combining data at $\sqrt{s}$=130--183~GeV. For $\sqrt{s}\leq 161$~GeV
%the efficiency is calculated as a function of $m_X/\sqrt{s}$, from that 
%obtained from Monte Carlo at $\sqrt{s}=183$~GeV.

The results obtained are valid for
particles with a lifetime longer than 10$^{-6}$~s.
This is calculated in the worst case, e.g. the heaviest (and 
therefore slowest) particles excluded by this search, 
requiring that the decay probability of these particles at a 
flight distance larger than 3.0~m is greater than 95\%. 
For lower mass values the results are also valid for shorter lifetimes.

%=======================================================================
\section{Summary and Conclusions}
%=======================================================================
\label{sec:conclusions}

A search has been performed for pair-production of stable and 
long-lived massive particles with 
charge $|Q/e|$~=~1 or $2/3$.
The primary tool used in this search is the precise $\dedx$ measurement 
provided by the OPAL jet chamber. 
No evidence for the production of heavy particles with masses
in the range of 45 to 89.5 GeV was observed. For s-channel production,
the upper limits on the cross-section 
vary between 0.05 and 0.19~pb in
the range of masses explored for particles of charge $\pm$1. 
Within the framework of the MSSM, lower mass limits on the 
the right- (left-) handed smuons and staus of
82.5~GeV (83.5~GeV) have been obtained. 
Heavy long-lived charged leptons and 
long-lived charginos with masses smaller than 89.5~GeV 
are excluded.
For particles with charge $\pm$2/3 the upper limits on the production
cross-section vary between 0.05 and 0.2~pb in the range of masses
explored.
The above limits are valid at the 95\% CL for particles with lifetimes
longer than 10$^{-6}$~s.

%=======================================================================
% OPAL acknowledgments
%=======================================================================
%% \input{opal_ack}
\bigskip\bigskip\bigskip
\appendix
\par
{\Large\bf Acknowledgements}
\par
We particularly wish to thank the SL Division for the efficient operation
of the LEP accelerator at all energies
 and for
their continuing close cooperation with
our experimental group.  We thank our colleagues from CEA, DAPNIA/SPP,
CE-Saclay for their efforts over the years on the time-of-flight and trigger
systems which we continue to use.  In addition to the support staff at our own
institutions we are pleased to acknowledge the  \\
Department of Energy, USA, \\
National Science Foundation, USA, \\
Particle Physics and Astronomy Research Council, UK, \\
Natural Sciences and Engineering Research Council, Canada, \\
Israel Science Foundation, administered by the Israel
Academy of Science and Humanities, \\
Minerva Gesellschaft, \\
Benoziyo Center for High Energy Physics,\\
Japanese Ministry of Education, Science and Culture (the
Monbusho) and a grant under the Monbusho International
Science Research Program,\\
German Israeli Bi-national Science Foundation (GIF), \\
Bundesministerium f\"ur Bildung, Wissenschaft,
Forschung und Technologie, Germany, \\
National Research Council of Canada, \\
Research Corporation, USA,\\
Hungarian Foundation for Scientific Research, OTKA T-016660, 
T023793 and OTKA F-023259.\\

%=======================================================================
%       References
%=======================================================================

%=======================================================================
%       Tables
%=======================================================================

%%%%%%%%%%%%%%%%%%%%%%%%%%%%%%%%%%%%%%%%%%%%%%%%%%%%%%%%%%%%%%%%%%%%%%%%
%%% \end{document}
%%%%%%%%%%%%%%%%%%%%%%%%%%%%%%%%%%%%%%%%%%%%%%%%%%%%%%%%%%%%%%%%%%%%%%%%

%=======================================================================
%       Figures
%=======================================================================
\newpage

%%%%\begin{figure}[htbp]
%%%%\centering
%%%%%%\begin{tabular}{cc}
%%%%\epsfig{file=limit_quark_new.eps,width=15cm}
%%%%%%\end{tabular}
%%%%\caption[]{\sl
%%%%  \protect{\parbox[t]{15cm}{
%%%%Model-independent 95\% CL upper limits on the pair-production
%%%%cross-section of spin-1/2 heavy charge-2/3 long-lived particles as a function
%%%%of their mass (solid line). In calculating the upper limit, the candidate
%%%%is considered as described in the text.
%%%%The vertical dashed line represents the kinematic limit.
%%%%} } } 
%%%%\label{fig:xsect-quark}
%%%%\end{figure}

%%%%%%%%%%%%%%%%%%%%%%%%%%%%%%%%%%%%%%%%%%%%%%%%%%%%%%%%%%%%%%%%%%%%%%%%
\end{document}